\documentclass[showpacs,aps,pra,superscriptaddress,10pt]{revtex4-1}

\usepackage{amsmath}
\usepackage{bm}
\usepackage{ucs}
\usepackage{geometry}
\usepackage{graphicx}
\usepackage[caption=false]{subfig}
\usepackage{epstopdf} 
\usepackage{refcount}
\usepackage{array}
\usepackage{hyperref}
\usepackage[dvipsnames]{xcolor}

\graphicspath{{./Figures/}}
\begin{document}
\title{Dark soliton collisions in superfluid Fermi gases}
\author{W. Van Alphen}
\email{wout.vanalphen@uantwerpen.be}
\affiliation{TQC, Universiteit Antwerpen, Universiteitsplein 1, B-2610 Antwerpen, Belgium}
\author{G. Lombardi}
\affiliation{TQC, Universiteit Antwerpen, Universiteitsplein 1, B-2610 Antwerpen, Belgium}
\author{S. N. Klimin}
\affiliation{TQC, Universiteit Antwerpen, Universiteitsplein 1, B-2610 Antwerpen, Belgium}
\affiliation{Department of Theoretical Physics, State University of Moldova, 2009 Chi\unichar{537}in\unichar{259}u, Moldova}
\author{J. Tempere}
\affiliation{TQC, Universiteit Antwerpen, Universiteitsplein 1, B-2610 Antwerpen, Belgium}
\affiliation{Lyman Laboratory of Physics, Harvard University, Cambridge, Massachusetts 02138, USA}
\begin{abstract} 
In this work dark soliton collisions in a one-dimensional superfluid Fermi gas are studied across the BEC-BCS crossover by means of a recently developed finite-temperature effective field theory [Eur. Phys. J. B {\bf{88}}, 122 (2015)]. The evolution of two counter-propagating solitons is simulated numerically based on the theory's nonlinear equation of motion for the pair field. The resulting collisions are observed to introduce a spatial shift into the trajectories of the solitons. The magnitude of this shift is calculated and studied in different conditions of temperature and spin-imbalance. When moving away from the BEC-regime, the collisions are found to become inelastic, emitting the lost energy in the form of small-amplitude density oscillations. This inelasticity is quantified and its behavior analyzed and compared to the results of other works. The dispersion relation of the density oscillations is calculated and is demonstrated to show a good agreement with the spectrum of collective excitations of the superfluid.
\end{abstract} 
\maketitle
\section{Introduction}
\label{sec:intro}
Solitons are among the most fascinating nonlinear phenomena in physics. Arising from an interplay between dispersive and nonlinear effects of the underlying medium, these solitary waves retain their shape while propagating at a constant velocity. Solitons emerge in a wide variety of physical systems, including optical fibers, classical fluids and plasmas. More recently, they have also become a subject of interest in the ultracold atoms community. Due to the fact that ultracold quantum gases are well-controlled and highly tunable nonlinear systems, they form an ideal environment for studying the properties and dynamics of solitons. In particular, solitons in ultracold atom clouds most often manifest themselves as \emph{dark} solitons, which are characterized by a localized density dip and a jump in the phase profile of the order parameter. Dark solitons have been theoretically and experimentally studied in Bose-Einstein condensates \cite{THFrantzeskakis,EXPDenschlag,EXPBurgerBongs,EXPAndersonHaljian,EXPBeckerStellmer} and superfluid Fermi gases \cite{THAntezzaDalfovo,THScottDalfovo,THLiaoBrand,EXPKu2}. 
In both of these systems, they are subject to an instability mechanism called the snake instability \cite{EXPKu2,THMuryshev,THBrandReinhardt,THCetoliBrand}, which makes the soliton decay into vortices if the radial width of the atom cloud is too large.
Since this inhibits the creation and observation of stable solitary waves in three-dimensional (3D) quantum gases, the preferred set-ups to study dark solitons are elongated quasi-one-dimensional (1D) clouds. \\
Solitons are often portrayed as particle-like excitations. The main reason for this is that they preserve their identity not only while propagating, but also when interacting with each other: when two solitons collide, they re-emerge again as two solitonic waves \cite{THZabuskyKruskal,THZabusky}. In the case of Bose-Einstein condensates, collisions between dark solitons have been thoroughly studied both theoretically \cite{THHuangVelardeMakarov,THBurgerCarr} and experimentally \cite{EXPStellmerBecker,EXPWellerRonzheimer}. For fermionic superfluids on the other hand, this topic has been investigated far less extensively. Dark soliton collisions have been theoretically analyzed in quasi-1D Fermi superfluids by means of a generalized nonlinear Schr\"{o}dinger equation \cite{THWenHuang} and through numerical simulations of the time-dependend Bogoliuobov-de Gennes (TDBdG) equations \cite{THScottDalfovo2}, but the methods that were used in these works imposed limitations on either the domain or the amount of the resulting data.
The goal of the present paper is to further extend the study of dark soliton collisions in superfluid Fermi gases by using a recently developed effective field theory (EFT) that is capable of describing Fermi superfluids across the BEC-BCS crossover regime in a wide temperature domain \cite{THKTLDEpjB}. This theory is based on the assumption that the order parameter changes slowly in both space and time, corresponding to the condition \citep{THGorkov} that the pair field should vary over a spatial region larger than the pair correlation length \citep{THPistolesiStrinati,THPalestiniStrinati}. The consequent limitations and validity of the EFT are discussed in Sec.\ \ref{ssec:val}. The theory has already been successfully employed in the description of both stable dark solitons and the snake instability mechanism in different regimes of temperature and population imbalance \cite{THKTDPrA,THLvAKTPrA,THLvAKTSI}. In this work, 
we use the EFT equation of motion that governs the dynamics of the order parameter to numerically simulate the collision of two solitons in a 1D Fermi superfluid and study the properties of the re-emerging solitons across the BEC-BCS crossover. We demonstrate how the collision introduces a phase shift into the space-time trajectories of the solitons and analyze the effects of temperature and spin-imbalance on this quantity. We observe that the soliton interactions become inelastic when moving away from the BEC-regime, the lost energy being converted into small-amplitude density ripples that emanate from the point of collision. We determine the dispersion of these ripples and compare it to the dispersion of the collective excitations of the superfluid. Wherever possible, we qualitatively compare our results and their validity to those of the works mentioned above. 
\\
The (quasi-)1D setting in which we investigate the collisions ensures that the solitons are stable with respect to the snake instability \citep{THLvAKTSI} and is indeed the regime of interest for most of the current theoretical and experimental studies on this subject. The 1D propagating soliton solutions which constitute the initial states for the numerical simulations are obtained from analytic expressions for the phase and amplitude profile of the order parameter, which were derived in \cite{THKTDPrA,THLvAKTPrA} for the case of a dark soliton on a uniform background. While traditionally ultracold gases are studied in set-ups with harmonic trapping potentials, the recent realization of box-like optical traps \cite{EXPGauntSchmidutz} provides an incentive to investigate uniform superfluids and the opportunity to experimentally test the predictions of the present work.\\
The remainder of the article is organized as follows: in Sec.\ \ref{sec:model} we give a brief overview of the theoretical model employed to examine ultracold fermionic systems. In Sec.\ \ref{sec:res}, we investigate and discuss the properties of dark soliton collisions in a 1D Fermi superfluid using numerical simulations. Finally, the conclusions of our study are presented in Sec.\ \ref{sec:concl}.

\section{Model \& method}
\label{sec:model}
\subsection{Effective field theory}
\label{ssec:EFT}
\begin{figure}[b]
\centering
\includegraphics[width=\textwidth]{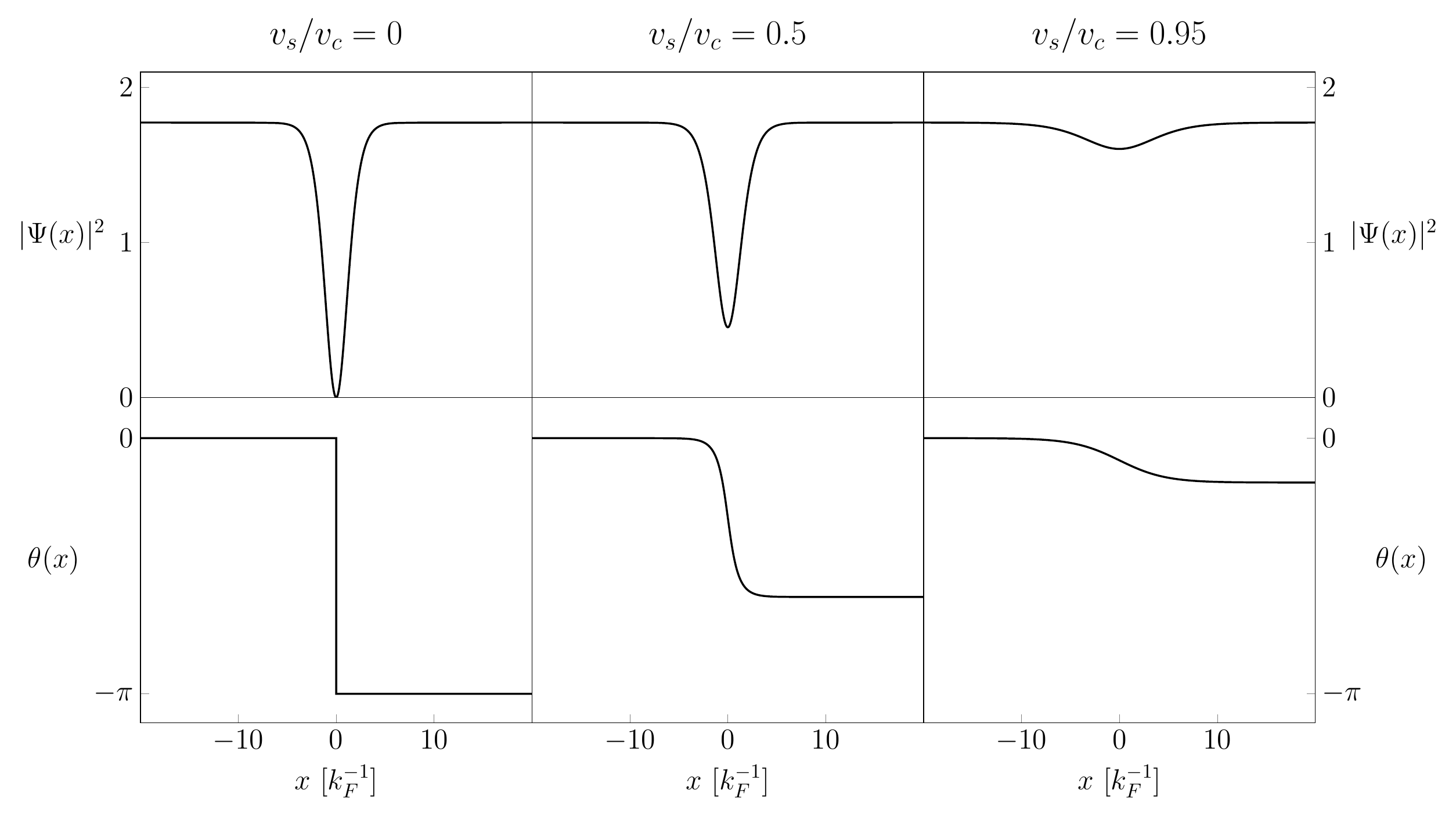}
\caption{Examples of the density profile (upper row) and phase profile (lower row) of a dark soliton for $(k_F a_s)^{-1} = 1$, for the velocity fractions $v_s/v_c = 0$ (left column), $v_s/v_c = 0.5$ (middle column) and $v_s/v_c = 0.95$ (right column). Here, $v_c$ is the critical velocity above which a soliton solution no longer exists. The position is given in units of $k_F^{-1}$.} 
\label{fig:solexamp}
\end{figure}

The system under consideration is an ultracold Fermi gas in which particles of opposite pseudo-spin interact via an $s$-wave contact potential. In the context of a recently developed finite-temperature effective field theory \cite{THKTLDEpjB}, this system can be described across the BEC-BCS crossover in terms of a superfluid order parameter $\Psi(\mathbf{r},t)$ (representing the bosonic pair field), under the assumption that $\Psi(\mathbf{r},t)$ varies slowly in both space and time. In the natural units of $\hbar = 1$, $2m = 1$, $E_F = 1$, the  Euclidean-time action functional for the system is then given by
\begin{equation}
\label{eq:action}
S[\Psi] = \int_0^{\beta} \mathrm{d}\tau \int \mathrm{d}\mathbf{r} \left[ \frac{D(\vert \Psi \vert )}{2} \left( \bar{\Psi} \frac{\partial \Psi}{\partial \tau} - \frac{\partial \bar{\Psi}}{\partial \tau} \Psi \right) + \mathcal{H} \right]
\end{equation}
where $\beta$ is the inverse temperature, and the Hamiltonian density $\mathcal{H}$ is given by
\begin{equation}
\mathcal{H} = \Omega_s( \vert \Psi \vert ) + \frac{C}{2 m} \vert \nabla_{\mathbf{r}} \Psi \vert^2 - \frac{E}{2m} \left( \nabla_{\mathbf{r}} \vert \Psi \vert^2 \right)^2
\end{equation}
The thermodynamic potential $\Omega_s$ reads
\begin{align}
\label{eq:omega}
\Omega_s(\vert \Psi \vert )=&-\int\frac{\mathrm{d}\bm{k}}{\left(2\pi\right)^3}\Bigg[\frac{1}{\beta}\log\left[2\cosh\left(\beta E_{\bm{k}}\right)+2\cosh\left(\beta \zeta\right)\right]+\nonumber\\&-\xi_{\bm{k}}-\frac{m\left|\Psi\right|^2}{k^2}\Bigg]-\frac{m\left|\Psi\right|^2}{4\pi a_s} 
\end{align}
while the coefficients $C$, $D$ and $E$ are defined as
\begin{align}
C &  =\int\frac{d\mathbf{k}}{\left(  2\pi\right)  ^{3}%
}\frac{k^{2}}{3m}f_{2}\left(  \beta,E_{\mathbf{k}},\zeta\right)  ,\label{eq:c}\\
D &  =\int\frac{d\mathbf{k}}{\left(  2\pi\right)  ^{3}}\frac
{\xi_{\mathbf{k}}}{w}\left[  f_{1}\left(  \beta,\xi_{\mathbf{k}},\zeta\right)
-f_{1}\left(  \beta,E_{\mathbf{k}},\zeta\right)  \right]  ,\label{eq:d}\\
E  &  =2 \int\frac{d\mathbf{k}}{\left(  2\pi\right)  ^{3}}%
\frac{k^{2}}{3m}\xi_{\mathbf{k}}^{2}~f_{4}\left(  \beta,E_{\mathbf{k}}%
,\zeta\right)  ,\label{eq:e}
\end{align}
The functions $f_j(\beta,\epsilon,\zeta)$ in the above expressions are defined by
\begin{align}
f_{j}(\beta,\epsilon,\zeta)=\frac{1}{\beta}\sum_{n}\frac{1}{\left[\left(\omega_n-i\zeta\right)^2+\epsilon^2\right]^j}
\end{align}
with the fermionic Matsubara frequencies $\omega_n=(2n+1)\pi/\beta$.
In this treatment, the chemical potentials of the two pseudo-spin species $\mu_{\uparrow}$ and $\mu_{\downarrow}$ are combined into the average chemical potential $\mu = (\mu_{\uparrow} + \mu_{\downarrow})/2$ and the  imbalance chemical potential $\zeta = (\mu_{\uparrow} - \mu_{\downarrow})/2$, the latter determining the difference between the number of particles in each spin-population. The quantity $\xi_{\mathbf{k}} = \frac{k^2}{2m} - \mu$ is the dispersion relation for a free fermion, $E_{\mathbf{k}} = (\xi_{\mathbf{k}} + \vert \Psi \vert^2)^{1/2}$ is the Bogoliubov excitation energy, and $a_s$ is the $s$-wave scattering length that determines the strength and the sign of the contact interaction. It is important to note that, while both the coefficients $C$ and $E$ in the action functional are kept constant and equal to the value they assume in the uniform system case, the coefficient $D$ and the thermodynamic potential $\Omega_s$ fully depend upon the order parameter \cite{THKTDPrA}. The regularized real-time Lagrangian density that follows from \eqref{eq:action} reads
\begin{equation}
\label{eq:rtlag}
\mathcal{L} = i\frac{D( \vert \Psi \vert )}{2} \left( \bar{\Psi} \frac{\partial \Psi}{\partial t} - \frac{ \partial \bar{\Psi}}{\partial t} \Psi \right) - (\mathcal{H} - \Omega_s(\vert \Psi_{\infty} \vert))
\end{equation}
where $\vert \Psi_{\infty} \vert$ is the value of the order parameter for a uniform system (i.e.\ the superfluid gap). Consequently, the subtraction of the term $\Omega_s(\vert \Psi_{\infty} \vert)$ means that in the following calculations all energy values are considered as energy differences with respect to the energy of the uniform system. Since this background value $\vert \Psi_{\infty}  \vert$ and the average chemical potential $\mu$ essentially serve as input parameters for the coefficients \eqref{eq:omega}-\eqref{eq:e}, there is a certain freedom of choice for the values of these quantities which allows to tune the quantitative accuracy of the EFT. In this work, we assign to these parameters the mean-field values that are obtained by simultaneously solving the saddle-point gap and number equations \cite{THDevreeseTempere}. At unitarity, this yields the values $\vert \Psi_{\infty} \vert = 0.69 \, E_F$, $\mu = 0.59 \, E_F$ and $T_c = 0.5 \, T_F$ for respectively the gap, chemical potential and critical temperature of the system. These background values could be further improved upon by, for example, including fluctuations around the saddle point, using the values of quantum Monte-Carlo simulations \cite{THCarlsonChang,THAstrakharchik,THCarlsonReddy} or even using the values derived from experimental measurements \cite{EXPKuSommer}. 
\\
From the Lagrangian density \eqref{eq:rtlag}, the equation of motion for the pair field $\Psi$ is found to be:
\begin{equation}
\label{eq:eqofmot}
i \tilde{D}(\vert \Psi \vert ) \frac{\partial \Psi}{\partial t} = -\frac{C}{2 m} \nabla^2_{\mathbf{r}} \Psi + \left( \mathcal{A}(\vert \Psi \vert ) + \frac{E}{m} \nabla^2_{\mathbf{r}} \vert \Psi \vert^2 \right) \Psi
\end{equation} 
where the coefficients $\tilde{D}$ and $\mathcal{A}$ are defined as
\begin{align}
\mathcal{A}  &  =\frac{\partial\Omega_{s} }{\partial \left( |\Psi|^2 \right)},\quad\tilde{D}
=\frac{\partial\left(  |\Psi|^2 D  \right)  }{\partial
\left(|\Psi|^2\right)}. \label{eq:A&D}
\end{align}
This equation is a type of non-linear Schr\"{o}dinger equation which is closely related to both the Gross-Pitaevskii (GP) equation for Bose-Einstein condensates and the Ginzburg-Landau (GL) equation for Fermi superfluids \citep{THRanderiaSaDeMelo}. The first term on the right-hand side can be identified as a kinetic energy term, while the non-linear term represents a system-inherent potential for the field. The ratio $\tilde{D}/C$ can be interpreted as a renormalization factor for the mass of the fermion pairs \cite{THKTVPrA94} and the coefficient $\mathcal{A}$ determines the uniform background value of the system, since $\mathcal{A}(\Psi) \, \Psi = 0$ is nothing but the aforementioned gap equation. It has been verified that in the deep BEC-limit $\left( 1/k_F a_S \gg 1 \right)$ this equation of motion correctly tends to the GP equation for bosons with a mass $M=2m$ and an s-wave boson-boson scattering length $a_B = 2 \, a_s$ \citep{THLombardiPhD}. In the most general case, equation \eqref{eq:eqofmot} describes a 3D fermionic superfluid. To limit ourselves to the study of (quasi-)1D systems with a uniform background, we will consider only solutions of the form $\Psi(\mathbf{r},t) = \Psi(x,t)$, thus imposing the one-dimensionality of the system directly onto the pair field. The 1D equation of motion then becomes:
\begin{equation}
\label{eq:eqofmot1D}
i \tilde{D}(\vert \Psi \vert ) \frac{\partial \Psi}{\partial t} = -\frac{C}{2 m} \frac{\partial^2 \Psi}{\partial x^2} + \left( \mathcal{A}(\vert \Psi \vert ) + \frac{E}{m} \frac{\partial^2 \vert \Psi \vert^2}{\partial x^2} \right) \Psi
\end{equation} 
The expressions for the coefficients $C$, $\tilde{D}$, $E$ and $\mathcal{A}$ remain the same as in \eqref{eq:omega}-\eqref{eq:e} and \eqref{eq:A&D}, but $\tilde{D}$ and $\mathcal{A}$ now only depend upon one spatial parameter through their dependence on $\Psi(x,t)$. For the case of a 1D dark soliton that propagates with a constant velocity $v_s$ on a uniform background, equation \eqref{eq:eqofmot1D} can be solved analytically and an exact solution $\Psi_s(x-v_s t)$ can be found \cite{THKTDPrA,THLvAKTPrA}. Specifically, we write the complex order parameter as $\Psi(x,t) = \vert \Psi_{\infty} \vert \, a(x,t) \, e^{i \theta(x,t)}$ and solve the resulting equations for the amplitude modulation $a(x)$ and the phase field $\theta(x)$. This results in an expression for the phase profile
\begin{equation}
\theta(x) = \frac{v_s \, m}{C} \int_{-\infty}^{x} \frac{D(a(x')) \, a^2(x') - D(a_{\infty})}{a^2(x')} \, dx'\label{eq:PhaseSolution}
\end{equation}
and a relation for the position (i.e.\ the distance from the soliton center) in function of the amplitude
\begin{equation}
x(a) = \pm \frac{\vert \Psi_{\infty} \vert}{\sqrt{2 m}}  \int_{a_0}^a \sqrt{\frac{C - 4 \, \vert \Psi_{\infty} \vert^2 \, a'^2 \, E}{X(a') - v_s^2 \, Y(a')}} da'
\label{AmplitudeSolution}
\end{equation}
where $X(a)$ and $Y(a)$ are defined as
\begin{align}
X(a) &= \Omega_s(a) - \Omega_s(a_{\infty}) \\
Y(a) &= m \vert \Psi_{\infty} \vert^2 \frac{(D(a) \, a^2 - D(a_{\infty}))^2}{2 \, C \, a^2}.
\end{align}
The relative bulk amplitude is given by $a_{\infty} = 1$, while the relative amplitude at the center of the soliton $a_0 = a(x = 0)$ is determined as the solution of $X(a_0) - v_s^2 \, Y(a_0) = 0$.
This solution for the pair field exhibits the characteristic phase jump and amplitude dip at the position of the dark soliton, both of which are intrinsically connected to the soliton velocity $v_s$: the lower the magnitude of $v_s$, the larger the phase jump across the soliton position and the deeper the density dip. A stationary soliton, which is often referred to as a \emph{black soliton}, possesses the highest possible phase jump $\vert \Delta \theta \vert = \pi$ and a maximal depth (zero pair density) at its center. Finite-velocity solitons, often called \emph{gray solitons}, will become less and less deep as their velocity increases, up until a critical velocity $v_c$ (which coincides with the sound velocity $c_s$ of the superfluid) for which both the density dip and phase jump disappear completely and a soliton solution no longer exists. This relation between the soliton velocity, depth and phase jump is illustrated in Figure \ref{fig:solexamp}. 
\\
To simulate a dark soliton collision in a 1D superfluid in the context of the above model, an initial state $\Psi_0(x,t)$ consisting of two counter-propagating solitons is numerically evolved in time, using the EFT equation of motion \eqref{eq:eqofmot}. This initial state is constructed by combining two analytical solutions $\Psi_s(x-x_1-v_1 t)$ and $\Psi_s(x-x_2-v_2 t)$ for dark solitons with respective velocities $v_1$ and $v_2$ and respective initial positions $x_1$ and $x_2$. While in general the superposition of two solutions of a nonlinear equation is not necessarily a solution of the equation as well, it has been demonstrated that, due to their localized character, the superposition principle holds asymptotically for solitary waves at large spatial separation \citep{THZakharovShabat}. Indeed, sufficiently far from the position of each soliton, the modulus and phase of the order parameter will assume constant values, so the two solutions can be connected to each other if the distance between $x_1$ and $x_2$ is sufficiently large.  
The subsequent numerical time evolution of the initial state is carried out by discretizing the space-time grid and applying a finite-difference fourth order Runge-Kutta (RK4) algorithm. A detailed explanation of this numerical procedure is given in Appendix \hyperref[sec:appA]{A}.

\subsection{Validity of the model}
\label{ssec:val}
There are in general two approximations that determine the validity and limitations of the effective field theory outlined above. The main assumption of the model is that the order parameter changes slowly in both space and time, which corresponds to the condition that the pair field should vary over a spatial region much larger than the pair correlation length (also referred to as the Pippard correlation length in the context of superconducting systems). A detailed study of the limitations imposed by this condition was carried out in \cite{THLvAKTPrA}. By comparing the soliton width to the pair correlation length for the whole relevant region of the $\{(k_F a_s)^{-1},\beta,\zeta\}$ space, it was revealed that the reliability of the theory is not guaranteed on the BCS-side of the resonance at low temperatures. Hence, results found in this regime should be treated with additional caution. A second remark that must be made is that in the full version of the effective field theory, as it was introduced in \cite{THKTLDEpjB}, the action functional \eqref{eq:action} contains additional terms that result in terms with second order time derivatives of the pair field in the equation of motion \eqref{eq:eqofmot} (with a form similar to the terms for the spatial derivatives). While it would in principle be better to use the full second order time derivative (SOTD) model, we found that numerical SOTD simulations of soliton-soliton collisions exhibit inherent instabilities in a large area of the parameter domain. These instabilities are most likely caused by the fast deformation of the condensate at the moment of the collision. We therefore omit the second order time derivatives and make use of the first order time derivative (FOTD) version of the theory. It has been verified that for low-velocity solitons these omitted terms are of less importance and the deviations between the SOTD and FOTD calculations are usually small \citep{THLombardiPhD}. 

\section{Results}
\label{sec:res}
In this section we present the results of the numerical simulations of dark soliton collisions in a 1D superfluid Fermi gas, based on the EFT model that was described in the previous section. The initial state of each collision process is constructed as described in Sec.\ \ref{ssec:EFT}. The initial positions of the counter-propagating solitons are always chosen as $x_1 = -20 \, k_F^{-1}$ and $x_2 = 20 \, k_F^{-1}$, while the box size is chosen to be $L = 300 \, k_F^{-1}$. The resolution of the spatial grid is taken to be $\Delta x = 0.02 \, k_F^{-1}$. Figure \ref{fig:stab} in Appendix \ref{sec:appA} indicates the corresponding maximal value the time step $\Delta t$ can have in each interaction regime for the simulation to remain stable.

\subsection{General collision process}
\begin{figure}[htbp]
\begin{minipage}{0.65\textwidth}
\centering
\subfloat[]{
\includegraphics[width=\linewidth]{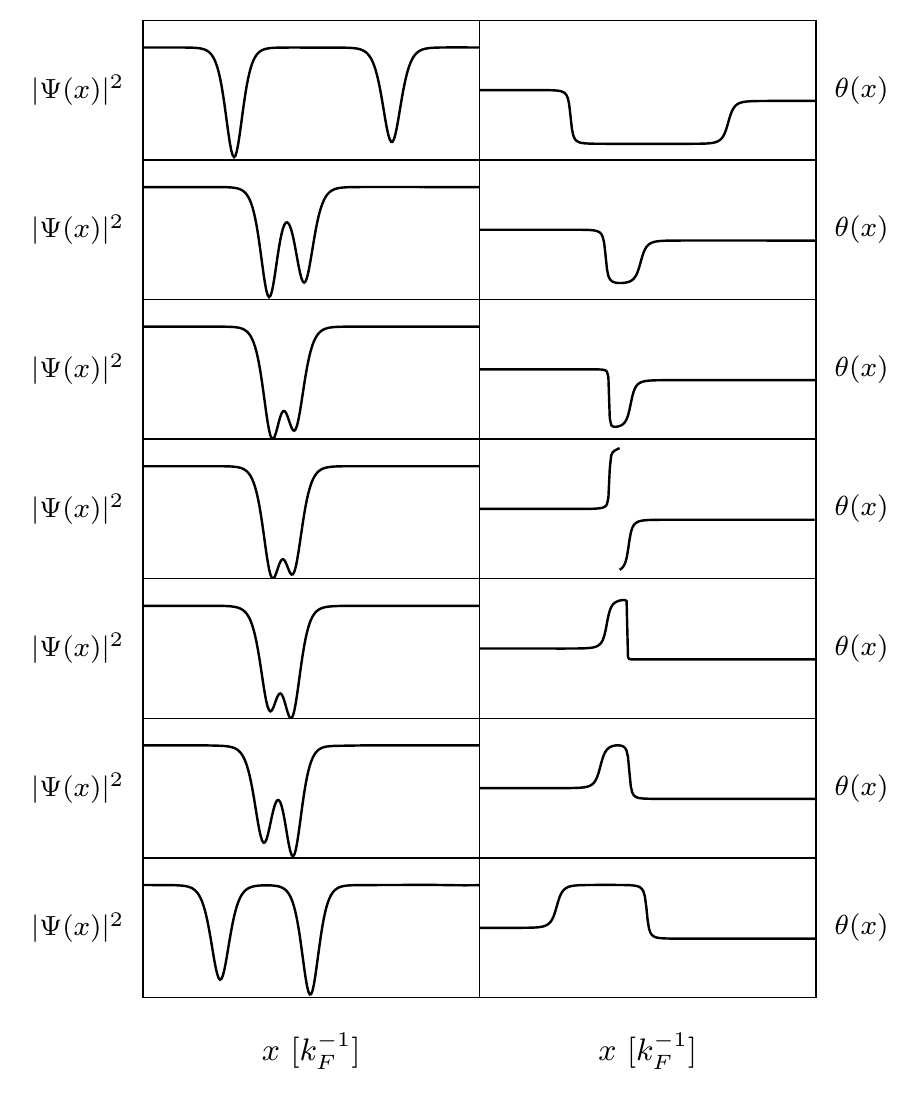}
\label{sfig:blackcolla}}
\end{minipage}
\begin{minipage}{.5\textwidth}
\centering
\subfloat[]{
\includegraphics[width=\linewidth]{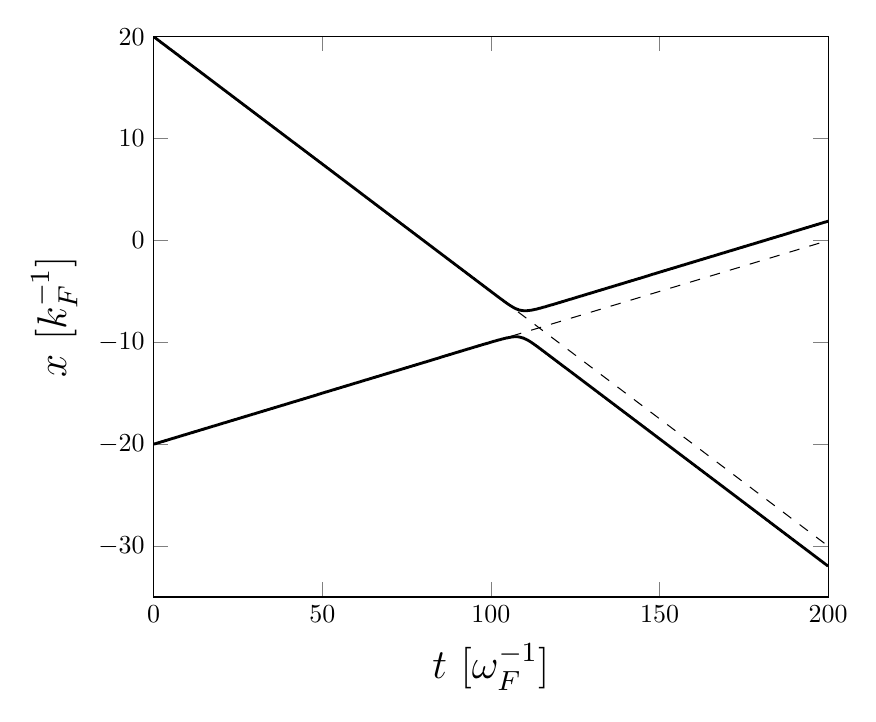}
\label{sfig:blackcollb}}
\end{minipage}
\caption{Example of the evolution of a black collision for $v_1 = 0.05 \, v_F$, $v_2 = 0.13 \, v_F$ and $(k_F a_s)^{-1} = 1$ (BEC-regime). Figure \ref{sfig:blackcolla} shows the evolution of the pair density (left panels) and the phase profile (right panels) of the superfluid for the successive time steps $t=50$, $t=100$, $t=106$, $t=108$, $t=111$, $t=115$ and $t=140$. Figure \ref{sfig:blackcollb} tracks the trajectories of the soliton centers. The dashed lines mark the trajectories each soliton would have followed if there was no collision. The position is given in units of $k_F^{-1}$ and the time in units of $\omega_F^{-1}$, with $\omega_F = E_F/\hbar$.} 
\label{fig:blackcoll}
\end{figure}
\begin{figure}[htbp]
\begin{minipage}{0.65\textwidth}
\centering
\subfloat[]{
\centerline{
\includegraphics[width=\linewidth]{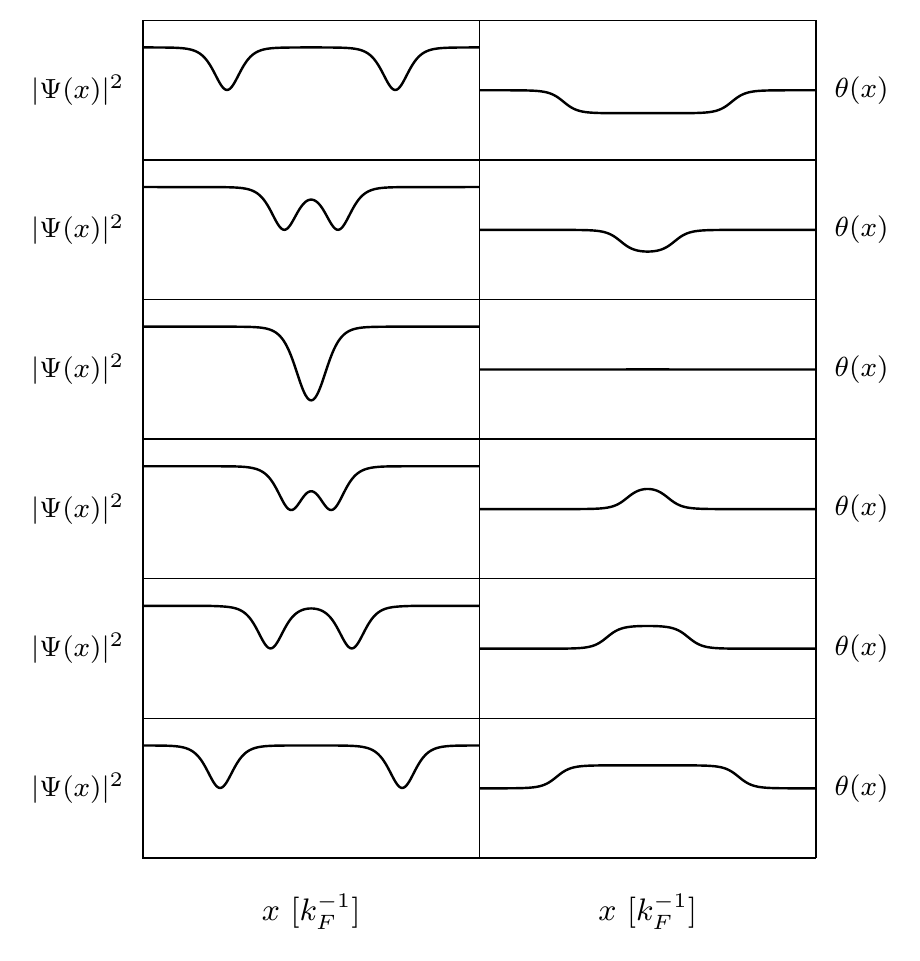}}
\label{sfig:graycolla}}
\end{minipage}
\begin{minipage}{.49\textwidth}
\centering
\subfloat[]{
\centerline{
\includegraphics[width=\linewidth]{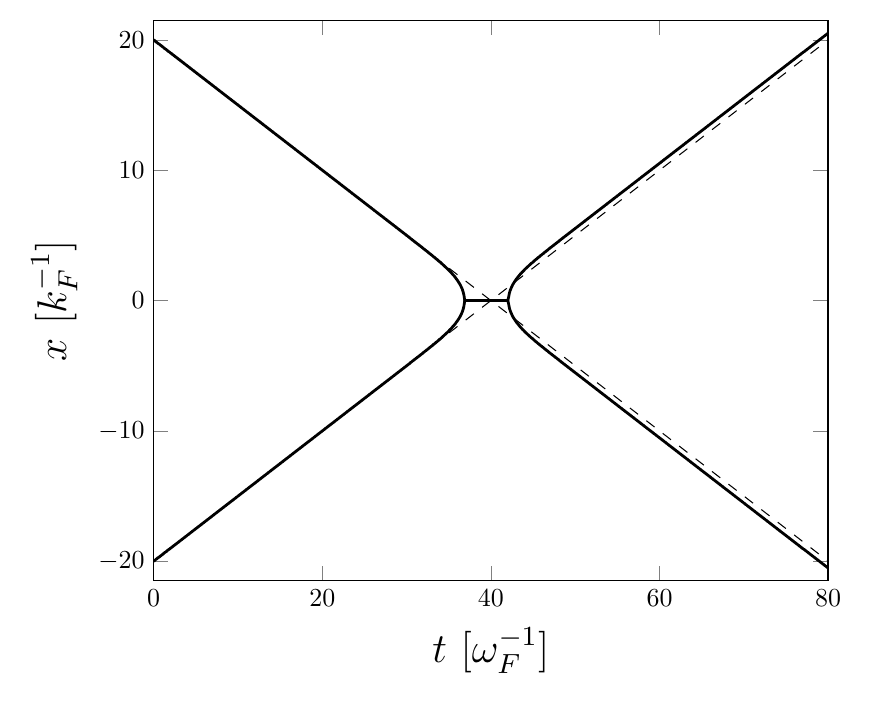}}
\label{sfig:graycollb}}
\end{minipage}
\caption{Example of the evolution of a gray collision for $v_1 = v_2 = 0.25 \, v_F$ and $(k_F a_s)^{-1} = 1$ (BEC-regime). Figure \ref{sfig:graycolla} shows the evolution of the pair density (left panels) and the phase profile (right panels) of the superfluid for the successive time steps $t=15$, $t=32$, $t=39$, $t=45$, $t=51$ and $t=66$. Figure \ref{sfig:graycollb} tracks the trajectories of the soliton centers. The dashed lines mark the trajectories each soliton would have followed if there was no collision. The position is given in units of $k_F^{-1}$ and the time in units of $\omega_F^{-1}$.} 
\label{fig:graycoll}
\end{figure}
We will start by describing and characterizing the evolution of the different types of dark soliton collisions. An important feature of these collisions is that the solitons preserve their identity, re-emerging as two solitary waves after the interaction. This is often interpreted as the fact that the two solitons simply propagate through each other. As an initial state, we consider two dark solitons with constant velocities $v_1$ and $v_2$, propagating towards each other through a uniform superfluid. 
Depending on the magnitude of these initial velocities, two different types of collisions can be identified. 
The first type of collision, which occurs if the initial velocities of the counter-propagating solitons are small enough, is illustrated in Figure \ref{fig:blackcoll}. The left panels of Figure \ref{sfig:blackcolla} show the evolution of the pair density profile of the superfluid, while the right panels show the evolution of the phase profile. We observe that at some time before the soliton centers reach other, both their respective velocities and the density at their centers decrease until they become zero.  Accordingly, the phase jump associated with each soliton grows until it reaches a value of $- \pi$, at which point it flips to a value of $\pi$ and causes the soliton velocity to change sign. As a result, the solitons propagate back into the opposite direction without having fully come together. On the space-time diagram \ref{sfig:blackcollb} that tracks the motion of the soliton centers, it appears as if the two solitons are simply reflected off each other in a particle-like collision. However, as can be observed from the evolution of the phase and density, the solitons actually exchange energy during the collision, adjusting their velocity, depth and phase jump accordingly. As a result, the soliton that was initially propagating to the right will still be found propagating to the right after the collision and vice versa, supporting the notion that the solitons move through one another. This type of collision is sometimes referred to as a \emph{black collision} \citep{THHuangVelardeMakarov}, because for each soliton there will be some point in time for which the density at its center vanishes completely. The second type of collision, which is sometimes referred to as a \emph{gray collision} \citep{THHuangVelardeMakarov}, is illustrated in a similar way in Figure \ref{fig:graycoll}. If the initial velocities of the solitons are sufficiently large, they will never become zero and the density at the soliton centers will never fully vanish during the collision. Instead, the solitons come together to form a single composite structure, at which point their phase jumps cancel each other out. After a limited amount of time, the solitons separate again and continue propagating through the superfluid, conforming to the picture that the solitons move through one another. It is interesting to relate this collision process to the general fact that creating a single density dip in a superfluid with a uniform phase background results in the creation of one or more pairs of finite-velocity solitons, a method that can be used for the density-engineering of solitons in experiments \cite{THBurgerCarr}. In the context of gray collisions, 
this is exactly what occurs during the recreation of the colliding solitons after their phases have canceled each other out. 
\subsection{Spatial shift of the soliton trajectory}
\begin{figure}[b]
\centering
\includegraphics[width=0.75\textwidth]{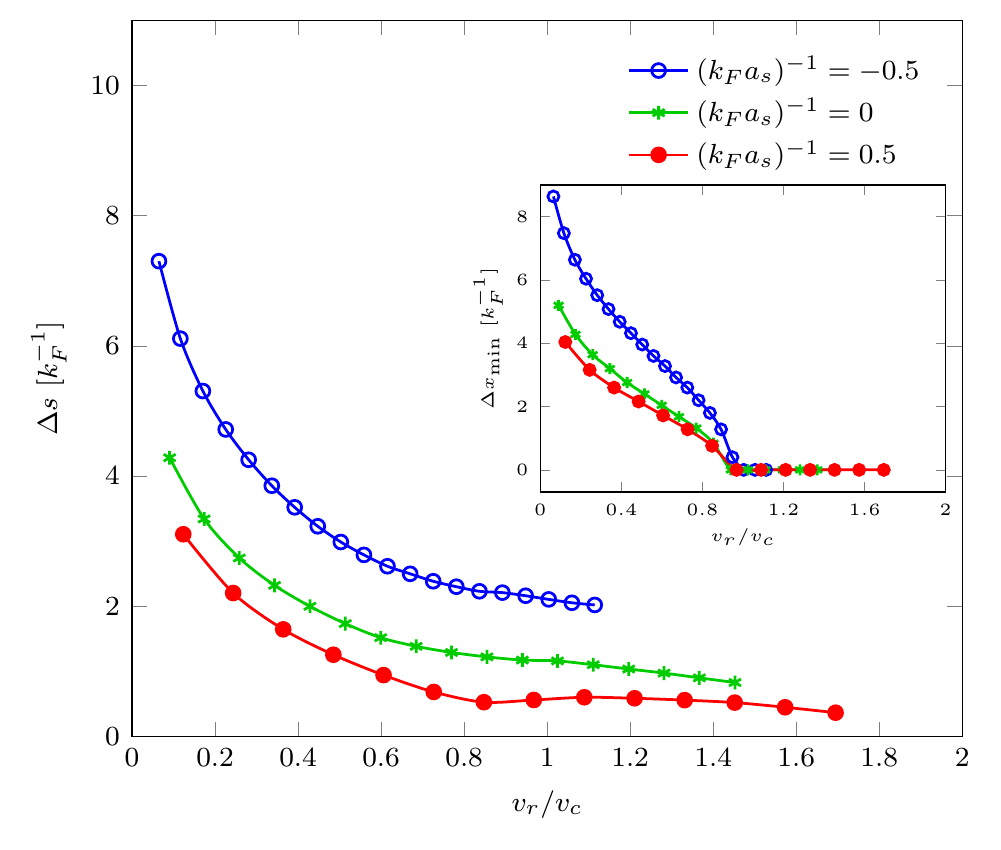}
\caption{The magnitude of the spatial shift $\Delta s$ for the symmetrical collision of two dark solitons in function of their relative velocity $v_r/v_c$ at temperature $T/T_F = 0.01$ for $(k_F a_s)^{-1} = 0.5$ (red full circles), $(k_F a_s)^{-1} = 0$ (green asterisks) and $(k_F a_s)^{-1} = -0.5$ (blue open circles). The FOTD values for the critical soliton velocity $v_c$ in these interaction regimes are respectively $v_c = 0.41 \, v_F$, $v_c = 0.59 \, v_F$ and $v_c = 0.90 \, v_F$. The inset shows the smallest occurring distance $\Delta x_{\text{min}}$ between the soliton centers in function of $v_r/v_c$. The quantities $\Delta s$ and $\Delta x_{\text{min}}$ are given in units of $k_F^{-1}$. The error bars on the data points are of the same size as the plot markers, so they are not explicitly depicted here.} 
\label{fig:shiftvs}
\end{figure}
Despite the notion that two interacting dark solitons simply propagate through each other, the collision does have observable effects on the outgoing solitons. The most important of these effects is that the post-interaction trajectories of the solitons acquire a \emph{positive} spatial shift with respect to the original trajectories. This can be clearly observed on the space-time diagrams in Figures \ref{sfig:graycolla} and \ref{sfig:graycollb}: after the collision both solitons have gained a positive shift in their own traveling directions when compared to their respective free paths (marked by the dashed lines). A consistent determination of the value of this spatial shift is complicated by the fact that the collisions become observably inelastic in certain velocity and interaction regimes, an effect which is further examined in Sec.\ \ref{ssec:inel}. Since in those cases the velocities of the outgoing solitons will have changed, the difference between the post-collision trajectory and the free trajectory will further increase as time goes on. For this reason, the independent effect of the spatial shift has to be studied sufficiently close to the collision time.
In order to maintain a clear graphical presentation, most results in this paper only involve symmetrical collisions, in which the solitons have opposite velocities of equal magnitude ($\vert v_1 \vert = \vert v_2 \vert$). In the more general case of asymmetrical collisions, the studied quantities depend on both of the individual soliton velocities instead of only the relative soliton velocity, but their general behavior still agrees with the results discussed below. 
\\ Figure \ref{fig:shiftvs} shows the magnitude of the spatial shift $\Delta s$ of the trajectory for a symmetrical collision of two dark solitons in function of their relative velocity $v_r/v_c$ for different values of the interaction parameter $(k_F a_s)^{-1}$ at temperature $T/T_F = 0.01$. 
As was mentioned in Sec.\ \ref{ssec:val}, the FOTD model of the EFT tends to exhibit deviations for high-velocity solitons and as a result overestimates the critical soliton velocity $v_c$ especially at the BCS-side of the interaction domain. To avoid going beyond the region of validity of our model, we refrain from showing results for soliton velocities above the values of $v_c$ predicted by the full SOTD formalism. From Figure \ref{fig:shiftvs},
it is clear that the shift of the trajectory becomes smaller as the relative velocity of the solitons increases, showing an initial rapid decrease and then leveling out slowly.
To explain this behavior, we also study the quantity $\Delta x_{\text{min}}$, which represents the smallest distance that occurs between the soliton centers during each collision. For a black collision, in which the solitons turn around before their centers reach each other, this value will be a finite number, while for a gray collision, in which the solitons merge completely, this quantity will be zero by definition. As can be seen in the inset of the figure, the behavior of $\Delta x_{\text{min}}$ in function of both $v_r$ and $(k_F a_s)^{-1} $ seems to be related to that of the curves in the main graph. Slow moving solitons with a large depth will reach their turning point while they are still at a relatively large distance from each other, and subsequently acquire a larger shift with respect to their free trajectory. When the solitons come closer to merging completely however, this effect diminishes
and the magnitude of the shift slowly levels out. 
The inset also indicates that the transition from a black to a gray collision consistently occurs when the relative velocity of the solitons is equal to the critical soliton velocity $v_c$. This was already demonstrated for the case of Bose-Einstein condensates \citep{TheocharisWeller}, but is now observed to be valid across the whole BEC-BCS crossover.  
\begin{figure}[tb]
\centering
\centerline{
\includegraphics[width=0.75\textwidth]{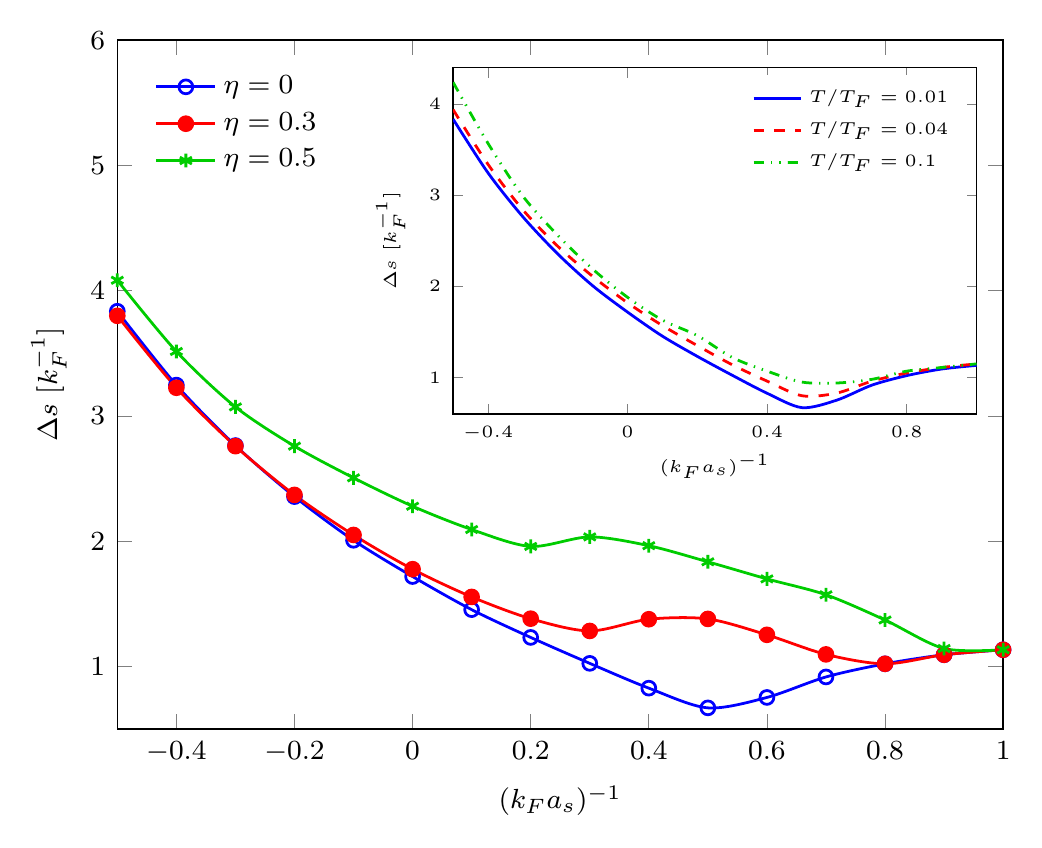}}
\caption{The magnitude of the spatial shift $\Delta s$ for the symmetrical collision of two dark solitons with relative velocity $v_r = 0.6 \, v_F$ in function of $(k_F a_s)^{-1}$ at $T/T_F = 0.01$ for $\eta = 0$ (blue open circles), $\eta= 0.3$ (red full circles) and $\eta = 0.5$ (green asterisks). 
The inset shows $\Delta s$ in function of $(k_F a_s)^{-1}$ for $T/T_F = 0.01$ (blue line), $T/T_F = 0.04$ (red dashed line) and $T/T_F = 0.1$ (green dotdashed line). The spatial shifts are given in units of $k_F^{-1}$. The error bars on the data points are of the same size as the plot markers, so they are not explicitly depicted here.} 
\label{fig:shiftka}
\end{figure}
\\
Figure \ref{fig:shiftka} shows the magnitude of $\Delta s$ for a symmetrical collision with relative velocity $v_r = 0.3 \, v_F$ across the BEC-BCS crossover for varying degrees of population-imbalance. The spin-imbalance parameter $\eta$ is defined as $\eta = \zeta/\vert \Psi_{\infty} \vert$ where $\vert \Psi_{\infty} \vert$ represents the superfluid gap of the homogeneous unpolarized gas \cite{THSonStephanov}. 
For the regular spin-balanced case ($\eta = 0$), $\Delta s$ decreases when moving away from the deep BEC-regime, reaches a minimum around $(k_F a_s)^{-1} = 0.5$ and then increases again towards the BCS-regime. The increase at the BCS-side of the resonance could be expected: the pair density at the center of the soliton is much lower in this interaction regime, causing the solitons to turn around sooner and thus acquiring a bigger shift. Additionally, the non-monotonous behavior of the shift around $(k_F a_s)^{-1} = 0.5$ can be explained by noticing that it is very similar to the behavior of the soliton width across the BEC-BCS crossover \cite{THLvAKTPrA}. This implies that solitons with a bigger width start feeling each other's influence sooner and as a result acquire a greater spatial shift in the collision. This idea also persists when we consider the effects of spin-imbalance on the value of the shift. It has been demonstrated that dark solitons broaden as the system becomes more imbalanced \cite{THLvAKTPrA}. In accordance with the above observation, this seems to result in an increase of the magnitude of the spatial shift with $\eta$. 
Using spin-imbalanced systems may therefore be a convenient way to make the spatial shifts of the solitons more clearly observable in experiments. The inset of the figure shows the effect of temperature on the shift. Since increasing the temperature slightly increases the soliton width, this also results in a small increase of $\Delta s$.  
\\
Our predictions for the behavior of $\Delta s$ across the BEC-BCS crossover in a spin-balanced Fermi superfluid can be qualitatively compared to those of Ref.\ \citep{THWenHuang}, where the authors also use a superfluid order parameter equation to calculate the relative shift of a collision between two high-speed solitons across the BEC-BCS crossover in a quasi-1D system. Their result displays a local minimum around $(k_F a_s)^{-1} = 0.5$, a local maximum close to unitarity and a subsequent decrease towards the deeper BCS-regime. While our own results also display a minimum around $(k_F a_s)^{-1} = 0.5$, we don't find a maximum and subsequent decrease of this quantity at the BCS-side of the interaction domain. In particular, since our observations suggest that the magnitude of the spatial shift depends on both the width and depth of the colliding solitons, and since neither of these quantities exhibits non-monotonous behavior in the BCS-regime, we do not find a reason for the spatial shift to display any non-monotonous behavior in this regime either.  
On the other hand, the approximations and calculations in Ref.\ \citep{THWenHuang} focus mainly on solitons with a high velocity, which happens to be a regime in which our own results might exhibit deviations on the BCS-side, as was discussed in Sec.\ \ref{ssec:val}. Consequentially, a definite conclusion can not be drawn from the available data.

\subsection{Inelasticity of the collisions}
\label{ssec:inel}
\begin{figure}[htbp]
\centering
\centerline{
\includegraphics[width=1.05\textwidth]{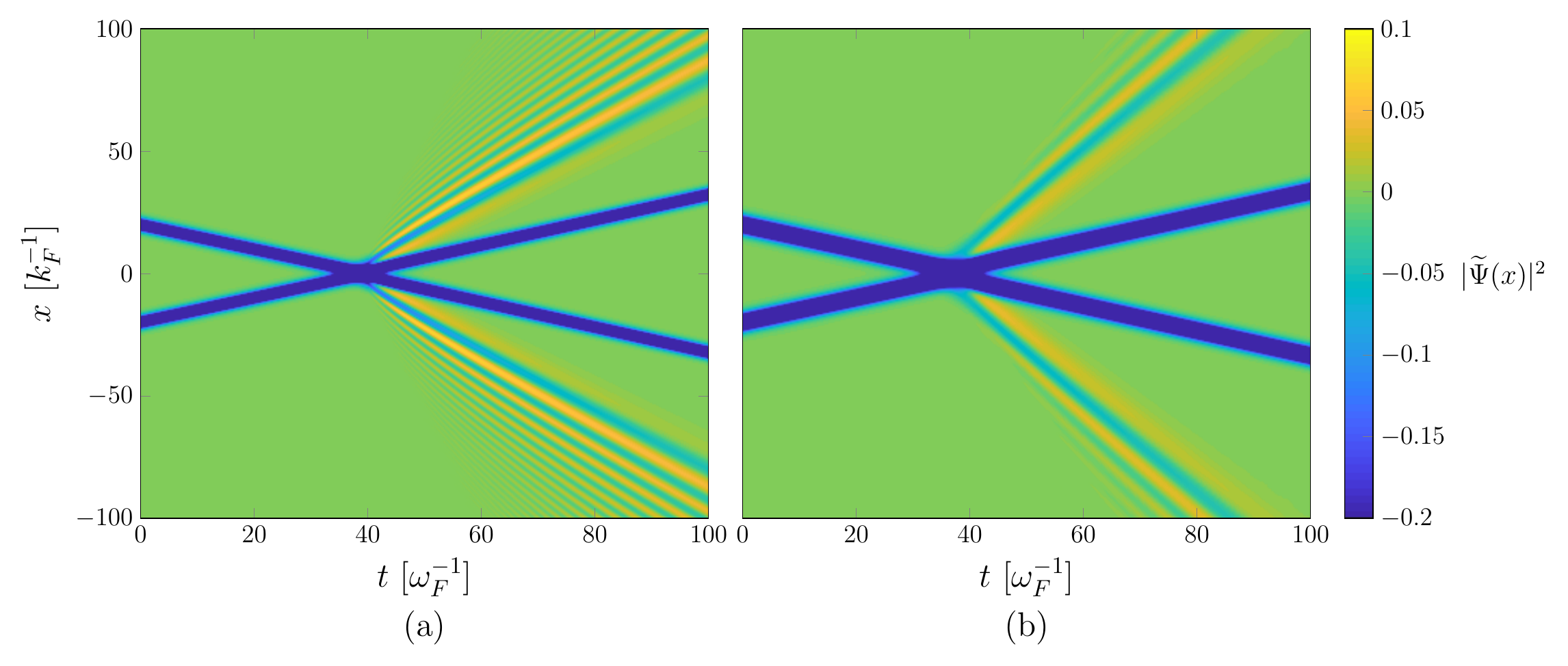}}
\caption{Space-time plot of the relative pair density profile $\vert \widetilde{\Psi} \vert^2 = \frac{\vert \Psi \vert^2 - \vert \Psi_{\infty} \vert^2}{\vert \Psi_{\infty} \vert^2}$  during a symmetric soliton collision with relative velocity $v_r = 0.5 \, v_F$ in (a) the unitarity regime ($(k_F a_s)^{-1} = 0$) and (b) the BCS-regime ($(k_F a_s)^{-1} = -0.5$). Blue colors indicate a lower relative density, yellow colors a higher relative density. One can clearly observe how density waves emanate from the point of collision, carrying away a small amount of energy from the colliding solitons. The position is given in units of $k_F^{-1}$ and the time in units of $\omega_F^{-1}$.} 
\label{fig:ripple}
\end{figure}
Our numerical simulations demonstrate that dark soliton collisions are not elastic across the whole BEC-BCS crossover. In the case of a (quasi-)1D BEC, soliton interactions are known to be completely elastic \citep{THHuangVelardeMakarov,EXPStellmerBecker}, and indeed we find that no inelasticity occurs in the deep BEC-limit of the interaction domain ($(k_F a_s)^{-1} \gg 1$). However, when moving away from the BEC-regime, collisions in certain velocity regimes become observably inelastic. The lost energy is converted into small-amplitude density ripples that emanate from the point of the collision, examples of which are shown in Figure \ref{fig:ripple} for both the unitarity and BCS-regime. Because solitons with a lower energy have a higher velocity and vice versa, the solitons actually move faster after losing energy in a collision and their inelasticity can be characterized by determining the velocity increase.
\begin{figure}[t]
\centering
\centerline{
\includegraphics[width=0.75\textwidth]{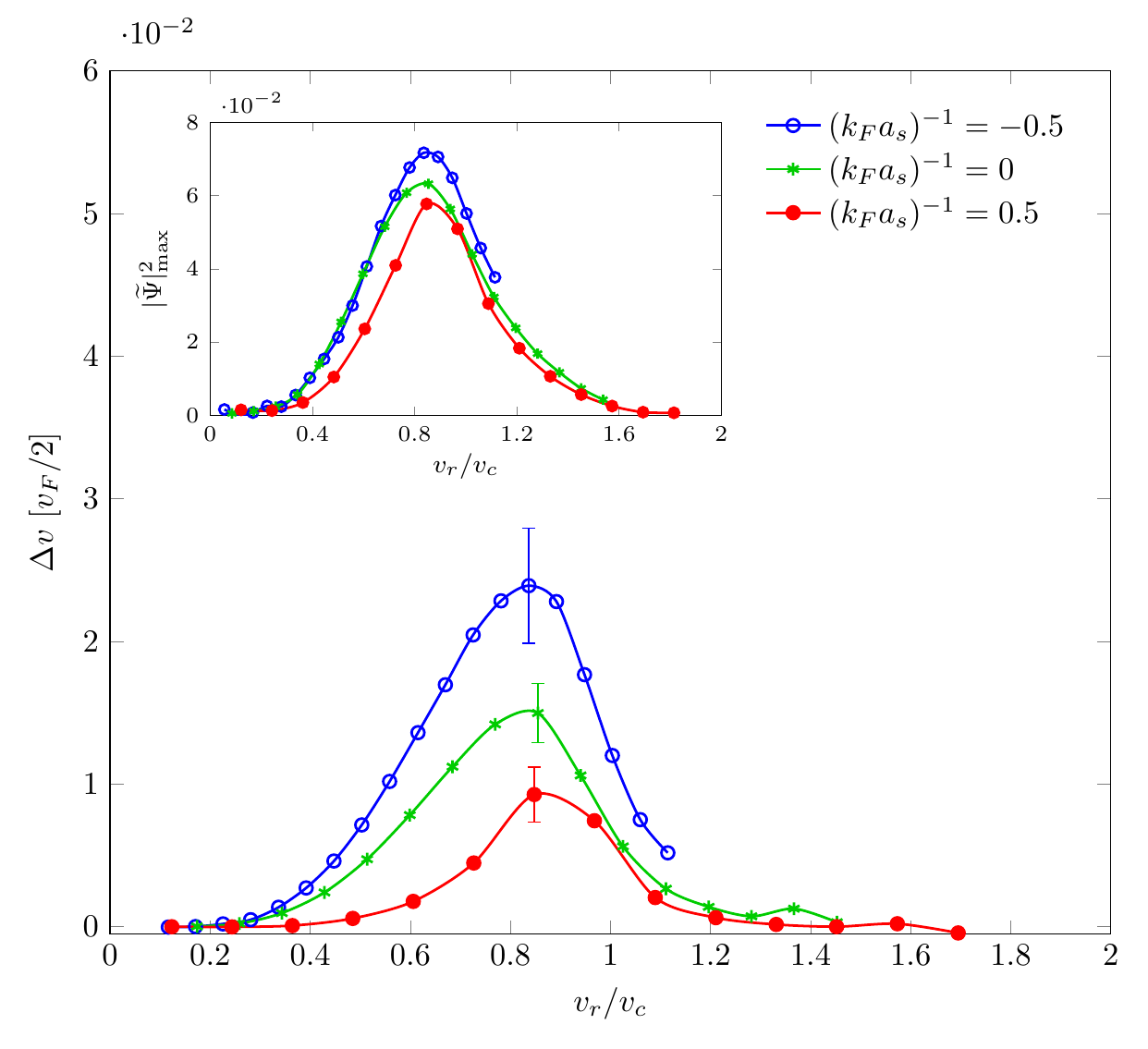}}
\caption{The magnitude of the velocity shift $\Delta v$ of a soliton after a symmetrical collision in function of the relative velocity $v_r/v_c$ at temperature $T/T_F = 0.01$ for $(k_F a_s)^{-1} = 0.5$ (red full circles), $(k_F a_s)^{-1} = 0$ (green asterisks) and $(k_F a_s)^{-1} = -0.5$ (blue clear circles). The FOTD values for the critical soliton velocity $v_c$ in these interaction regimes are respectively $v_c = 0.41 \, v_F$, $v_c = 0.59 \, v_F$ and $v_c = 0.90 \, v_F$. The inset shows the largest occurring pair density value $\vert \tilde{\Psi} \vert^2_{\text{max}}$ in function of $v_r/v_c$. 
The velocity differences are given in units of $v_F/2$. To depict the results in a clear way, we show the error bars for just one of the points of each data curve.} 
\label{fig:inelastvs}
\end{figure}
Figure \ref{fig:inelastvs} shows the magnitude of the velocity change $\Delta v$ of an individual soliton after a symmetrical collision, in function of the relative soliton velocity $v_r/v_c$ for different values of the interaction parameter $(k_F a_s)^{-1}$. The inset shows the maximum value assumed by the relative pair density $\vert \widetilde{\Psi} \vert^2_{\text{max}} = \frac{\vert \Psi \vert^2_{\text{max}} - \vert \Psi_{\infty} \vert^2}{\vert \Psi_{\infty} \vert^2}$ which identifies the largest density wave that is produced in each of these collisions. One can clearly see the correlation between the behavior of these quantities: the larger the generated density waves are, the more energy the solitons lose in the collision, and the higher the resulting velocity increase. In general, the inelasticity is small for slow-moving solitons, then increases to reach a maximum value at a fixed value of $v_r/v_c$, and finally becomes very small again for high-velocity solitons. Considering how the critical soliton velocity and thus the velocity range become larger when moving towards the BCS-regime, the inelasticity will peak at increasingly higher velocities.
\begin{figure}[t]
\centering
\centerline{
\includegraphics[width=0.75\textwidth]{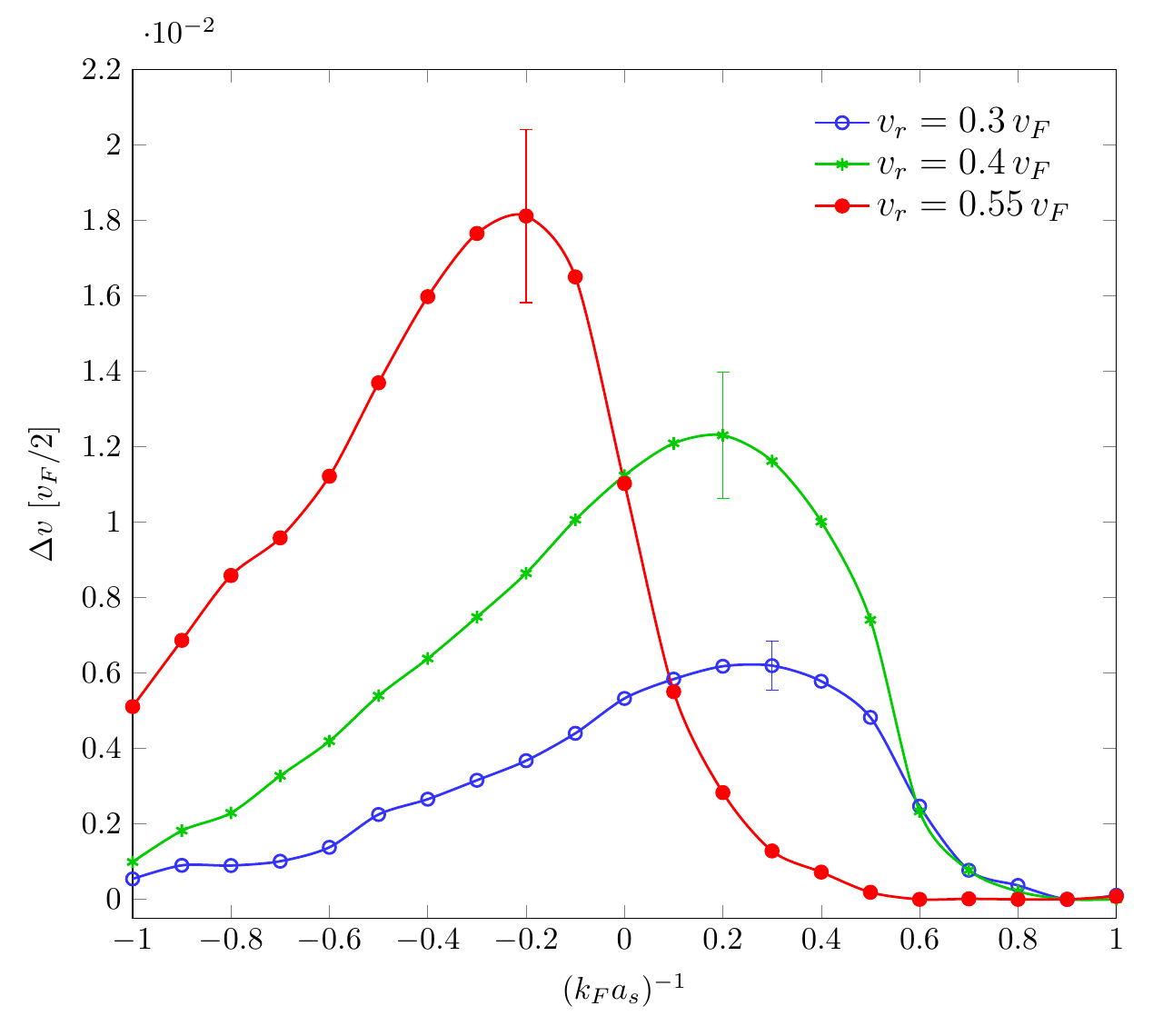}}
\caption{The magnitude of the velocity shift $\Delta v$ of a soliton after a symmetrical collision in function of the interaction parameter $(k_F a_s)^{-1}$ at temperature $T/T_F = 0.01$ for relative velocities $v_r = 0.3 \, v_F$ (blue open circles), $v_r = 0.4 \, v_F$ (green asterisks) and $v_r = 0.55 \, v_F$ (red full circles). The velocity differences are given in units of $v_F/2$. To depict the results in a clear way, we show the error bars for just one of the points of each data curve.} 
\label{fig:inelastka}
\end{figure}
Figure \ref{fig:inelastka} shows the magnitude of the velocity change $\Delta v$ of each soliton across the BEC-BCS crossover for different values of the relative velocity $v_r$.  The position of the maximum value of the curves indeed shifts towards the BCS-regime as the relative velocity increases. In the deep BEC-regime the inelasticity always becomes negligible, consistent with the fact that soliton collisions in (quasi-)1D BECs should be perfectly elastic. Moreover, we observe a similar downward trend for $\Delta v$ towards the other side of the interaction domain, implying that also in the deep BCS-regime the collisions become elastic again.
\\
Our results for the inelasticity of soliton collisions across the BEC-BCS crossover can be compared to those of Ref.\ \citep{THWenHuang} and \citep{THScottDalfovo2}. In the former work, in which the authors also make use of a nonlinear macroscopic order parameter equation, soliton collisions are observed to be as good as elastic across the whole BEC-BCS crossover. In the latter work on the other hand, in which a harmonically trapped Fermi superfluid is studied through numerical simulations of the TDBdG equations, inelastic soliton collisions are clearly observed. 
Due to the fact that the TDBdG method is computationally very demanding, only four values of $\Delta v$ were calculated across the BEC-BCS crossover. While this prevents a rigorous comparison between the different formalisms, we can nevertheless observe that the velocity increases found with the TDBdG method are almost one order of magnitude larger than those found in the present work. The authors determine that the inelasticity is caused by fermionic quasiparticles that are localized in the solitons (i.e.\ Andreev bound states) and identify this as the reason that in Ref.\ \citep{THWenHuang} only elastic collisions were observed, since these type of localized quasiparticle states are not present in a bosonic effective field theory. In the context of the currently applied EFT, the existence of a continuum of fermionic single-particle excitations with energy $E_{\mathbf{k}}$ does enter the formalism in an implicit way through the background theory, but no dynamical pair-breaking processes are contained within the Lagrangian for the bosonic pair field. This means that, similar to Ref.\ \citep{THWenHuang}, localized quasiparticles and Andreev bound states are absent in the current description. However, the present results do reveal a small but observable inelasticity for the collisions. This suggests that the reason that no inelastic collisions were observed in Ref.\ \citep{THWenHuang} might actually be due to the fact that the work focuses on solitons with a very high velocity, for which, according to the results in Figure \ref{fig:inelastvs}, the energy loss will become close to negligible. On the other hand, the fact that the observed energy losses are much larger in the TDBdG formalism than in the present case indicates that, while the inelasticity might not be completely due to the localized fermionic quasiparticles, they do provide a large contribution to this process.

\subsection{Dispersion of the density oscillations}
\begin{figure}[hbtp]
\begin{minipage}{0.49\textwidth}
\centering
\subfloat[]{
\includegraphics[width=\linewidth]{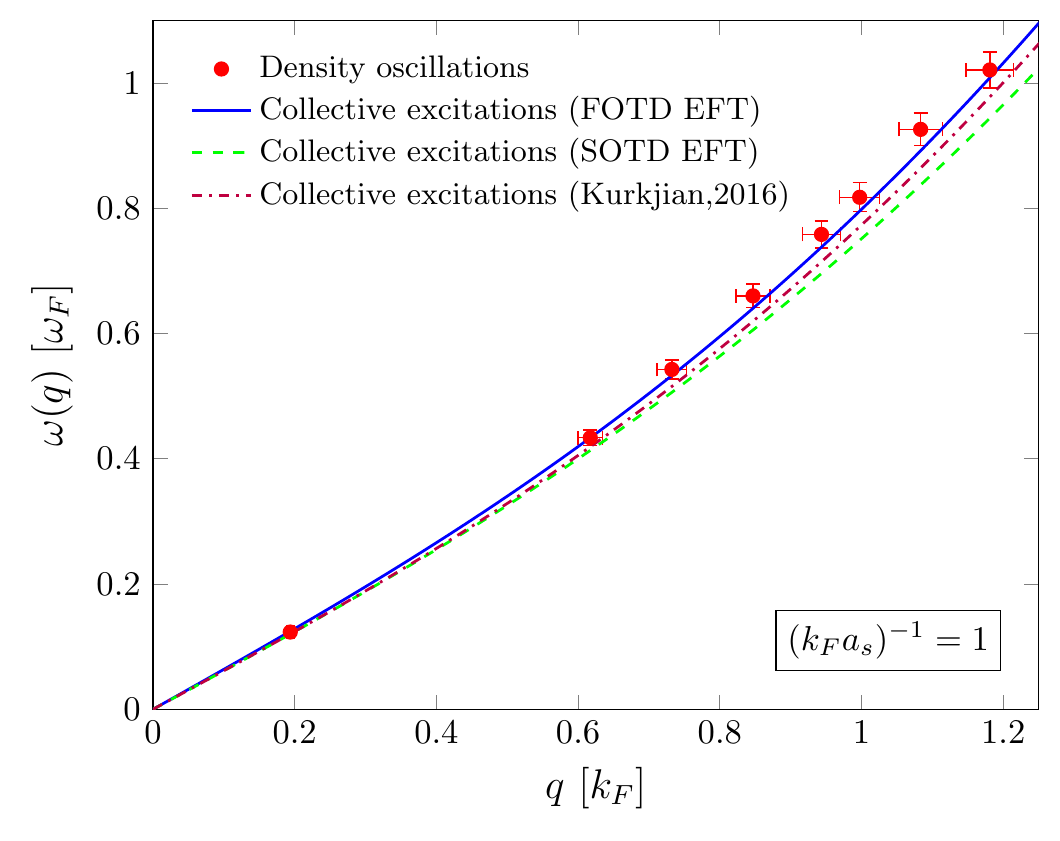}
\label{sfig:ripdispka1}}
\end{minipage}
\begin{minipage}{.49\textwidth}
\centering
\subfloat[]{
\includegraphics[width=\linewidth]{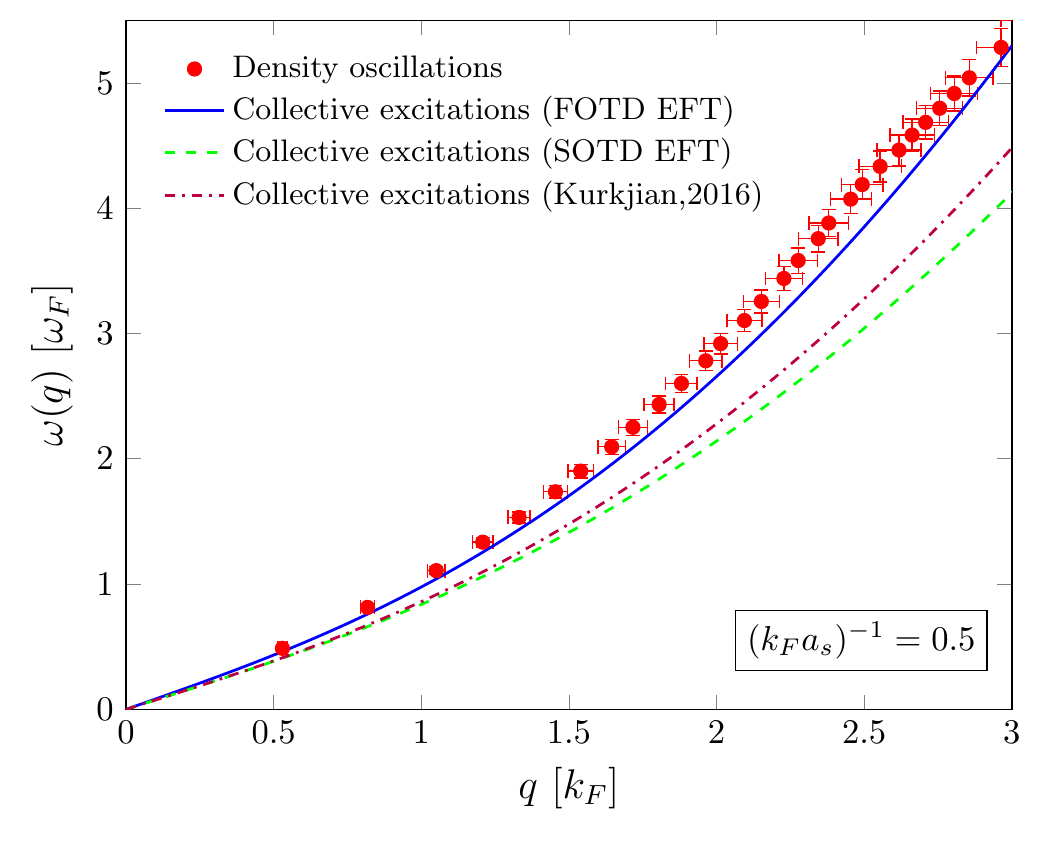}
\label{sfig:ripdispka05}}
\end{minipage}
\begin{minipage}{.49\textwidth}
\centering
\subfloat[]{
\includegraphics[width=\linewidth]{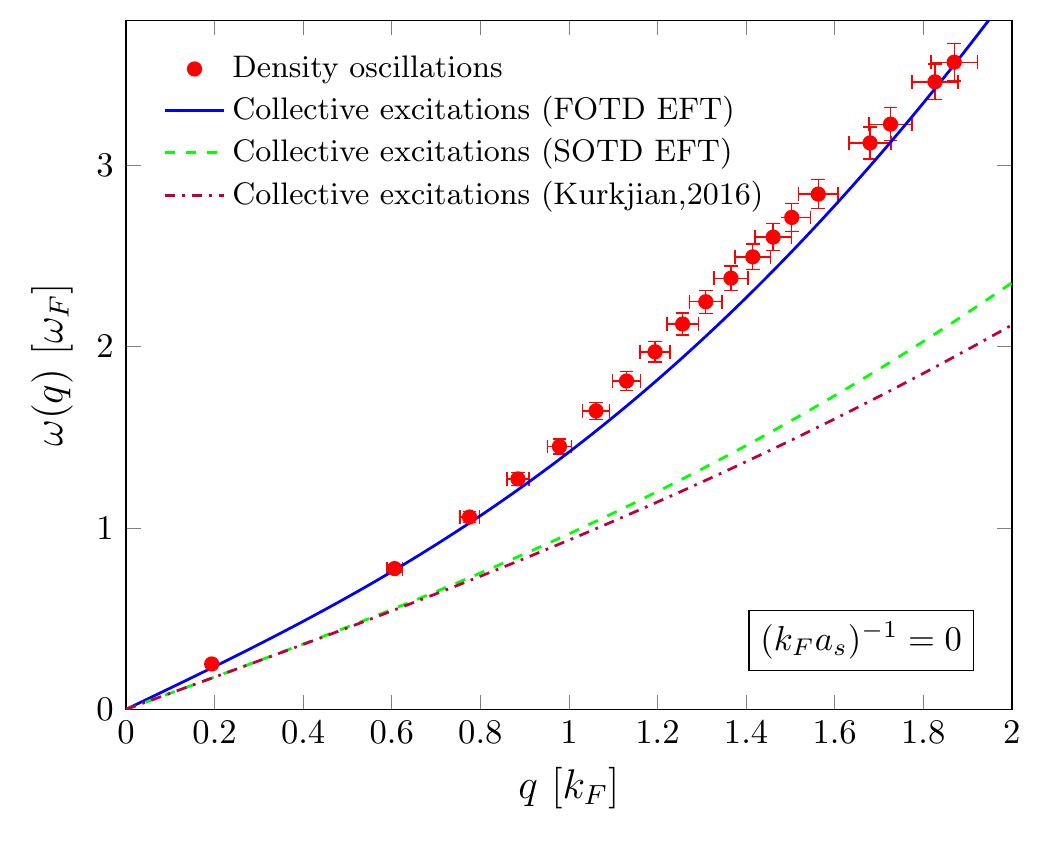}
\label{sfig:ripdispka0}}
\end{minipage}
\begin{minipage}{.49\textwidth}
\centering
\subfloat[]{
\includegraphics[width=\linewidth]{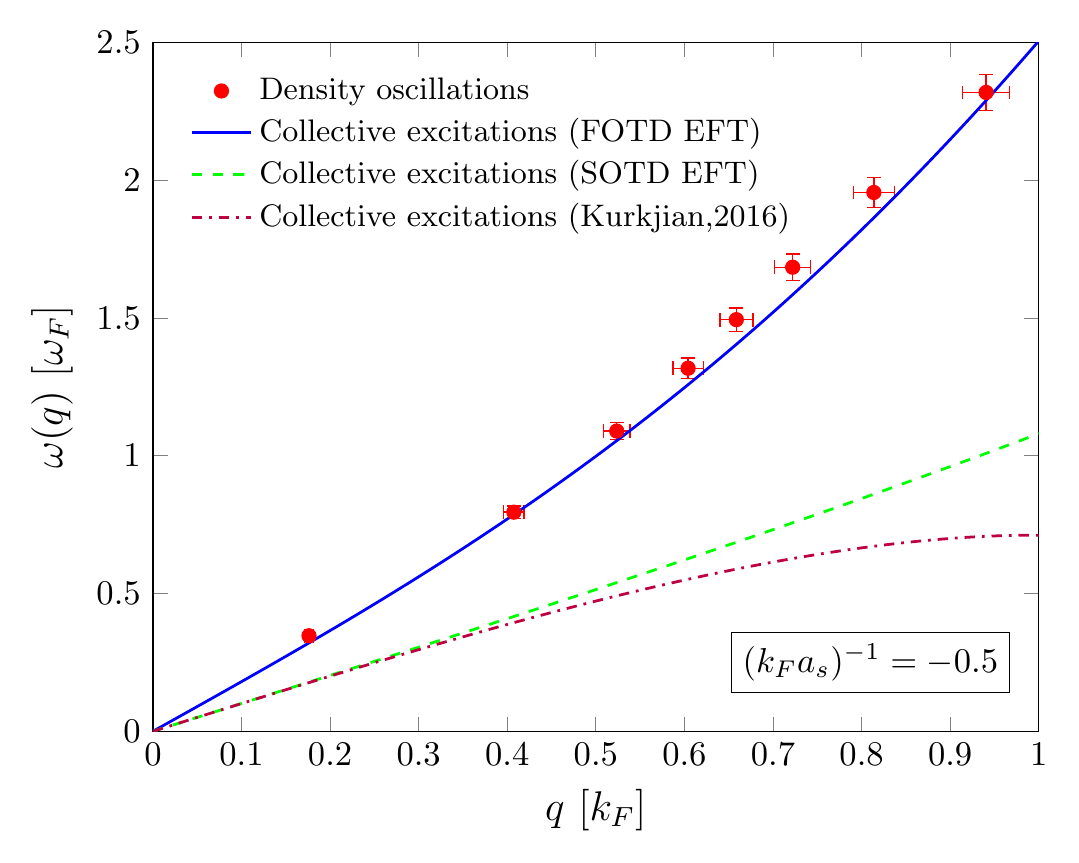}
\label{sfig:ripdispkam05}}
\end{minipage}
\caption{The dispersion of the density oscillations (red full circles) is compared to the spectrum of collective excitations for (a) $(k_F a_s)^{-1} = 1$, (b) $(k_F a_s)^{-1} = 0.5$, (c) $(k_F a_s)^{-1} = 0$ and (d) $(k_F a_s)^{-1} = -0.5$. For the spectrum of collective excitations, we show the EFT predictions of both the first order time derivative (blue full line) and second order time derivative (green dashed line) version of the theory, as well as the predictions of Ref.\ \citep{THKurkjian} (purple dash-dotted line). The wavenumbers are given in units of $k_F$ and the frequencies in units of $\omega_F$.} 
\label{fig:ripdisp}
\end{figure}
As can be observed in the space-time plot of the pair density profile in Figure \ref{fig:ripple}, the density ripples that emerge from an inelastic soliton collision generate a \emph{chirp}-like wave pattern in which the frequency of the oscillations increases along the length of the wave train. In order to obtain the corresponding dispersion relation $\omega(q)$, we determine the wavelength $\lambda_i$ (measured as the distance between two consecutive peaks) and the velocity $v_i$ of each matter wave and apply the relations $q_i = 2 \pi / \lambda_i$ and $\omega_i = q_i v_i$. The resulting dispersion of the oscillations can then be compared to the spectrum $\omega_s(q)$  of the collective excitations of the superfluid, which in the context of the present EFT is determined by calculating the poles of the inverse propagator for the bosonic fluctuations \citep{THKTLDEpjB}. Up to fourth order in $q$, this leads to the form
\begin{equation}
\omega_s^2(q) = c_s^2 q^2 + \lambda \left( \frac{q^2}{2m} \right)^2
\end{equation}
where the sound velocity $c_s$ and the coefficient $\lambda$ depend on the system parameters. While the EFT values for $c_s$ at zero temperature show a very good agreement with the results of Gaussian pair fluctuation calculations across the whole BEC-BCS crossover \citep{THKTLDEpjB}, the predictions for the coefficient $\lambda$ show deviations from those found by \citeauthor{THKurkjian} \citep{THKurkjian} when moving towards the BCS-regime \citep{THLombardiPhD}. It is also important to recall that, in the context of the current work, we have omitted from the EFT equation of motion all second order time derivatives of the pair field, limiting ourselves to only first order temporal variations. As was noted in Sec.\ \ref{ssec:val}, this introduces deviations in particular for high-velocity excitations and on the BCS-side of the resonance. In Figure \ref{fig:ripdisp}, the dispersion of the density ripples is compared to the spectrum of collective excitations of the superfluid for several values of the interaction parameter $(k_F a_s)^{-1}$ at near-zero temperature. For the collective excitation spectrum, we show the EFT results of both the FOTD and SOTD versions of the theory, as well as the results of Ref.\ \citep{THKurkjian}. As expected, the difference between the FOTD and SOTD curves becomes larger both at higher $q$-values and when moving towards the BCS-regime. Within the context of the FOTD calculations, the dispersion of the density oscillations shows a good agreement with the spectrum of collective excitation modes, identifying the wave trains as regular bosonic excitations of the superfluid. The slope of the ripple dispersion seems to tend correctly to the value of the sound velocity $c_s$ at low $q$. This agreement between the excitation spectrum and density wave dispersion is also observed at higher values of the temperature.

\section{Conclusions}
\label{sec:concl}
In this paper we studied the collisions of dark solitons in 1D superfluid Fermi gases by means of a finite-temperature effective field theory already employed to describe the properties of single dark solitons. Using numerical simulations based on the EFT equation of motion, we demonstrated that the collisions introduce a spatial shift into the soliton trajectories and we studied the effects of spin-imbalance and temperature on its magnitude. The fact that the presence of spin-imbalance increases the magnitude of the spatial shift could provide a convenient way to make this quantity easier to observe in experiments. When moving away from the deep BEC-regime, soliton collisions were found to become inelastic, resulting in an increase of the solitons' velocities after the collision. The inelasticity peaks in the middle of the soliton velocity range, but is close to negligible for very slow and very fast moving solitons. This could explain the absence of inelastic collisions in Ref.\ \citep{THWenHuang}, where only high-velocity solitons were studied. The fact that the observed changes in velocity are much smaller than in Ref.\ \citep{THScottDalfovo2} corroborates the fact that the main contribution to the inelasticity in Ref.\ \citep{THScottDalfovo2} is due to the presence of localized fermionic quasi-particles, which are absent in the current theory. The energy that is lost in the inelastic collisions was observed to be converted into trains of small-amplitude density oscillations, whose dispersion was demonstrated to show a good agreement with the spectrum of collective excitations of the superfluid.  
The numerical simulations carried out in this work were based on analytical solutions for 1D stable solitons in a uniform superfluid, as derived in \cite{THKTDPrA,THLvAKTPrA}. Even though ultracold gases are traditionally studied in set-ups with harmonic trapping potentials, the recent realization of box-like optical traps \cite{EXPGauntSchmidutz} can provide the opportunity to test the predictions of this work in experiment. While the effects of a single soliton collision is most likely too small to be reliably measured, the (necessarily) non-uniform boundaries of the trapping potential can be used to make the solitons turn around and undergo multiple repeated collisions in the uniform region. Subsequently, the accumulated effect of these consecutive collisions might be observable in experiments. 

\acknowledgements The authors acknowledge fruitful discussions with H.\ Kurkjian, N.\ Verhelst, S.\ Van Loon and T.\ Ichmoukhamedov. W.\ Van Alphen acknowledges financial support in the form of a Ph.\ D.\ fellowship of the Research Foundation - Flanders (FWO). This research was supported by the University Research Fund (BOF) of the University of
Antwerp and by the Flemish Research Foundation (FWO-Vl), project nr G.0429.15.N.

\appendix
\section{Discretization and evolution of the equation of motion}
\label{sec:appA}
\def\theequation{A.\arabic{equation}}
\setcounter{equation}{0}
\def\thesubsection{A.\arabic{subsection}}
\subsection{Finite-difference algorithm}
In this appendix we elaborate on how equation \eqref{eq:eqofmot1D} is discretized and solved numerically using the explicit RK4 algorithm. We begin by writing the equation of motion as
\begin{equation}
\frac{\partial \Psi}{\partial t} = f(x,t)
\end{equation}
with
\begin{equation}
\label{eq:f}
f(x,t) = i \frac{C}{2 m \tilde{D}[\Psi(x,t)]} \frac{\partial^2 \Psi(x,t)}{\partial x^2} - i \frac{\Psi(x,t)}{\tilde{D}[\Psi(x,t)]} \left( \mathcal{A}[\Psi(x,t)] + \frac{E}{m} \frac{\partial^2 \vert \Psi(x,t) \vert^2}{\partial x^2} \right)
\end{equation}
Square brackets are used to more clearly indicate the dependence of $\tilde{D}$ and $\mathcal{A}$ on $\Psi(x,t)$. Using a finite mesh width $\Delta x$ and a finite time step $\Delta t$, we discretize space-time into a grid of $M \times N$ points by writing $x_m = m \Delta x$ with $m = 1,...,M$ and  $t_n = n \Delta t $ with $n = 1,...,N$. This allows us to approximate the spatial derivatives in \eqref{eq:f} by central finite difference formulas. Using the notations $\Psi_{m,n} = \Psi(x_m,t_n)$, $w_{m,n} = \vert \Psi(x_m,t_n) \vert^2$, $\tilde{D}_{m,n} = \tilde{D}[\Psi(x_m,t_n)]$ and $\mathcal{A}_{m,n} = \mathcal{A}[\Psi(x_m,t_n)]$, the discretized form of \eqref{eq:f} is given by
\begin{equation}
f(\Psi_{m,n}) = i\frac{C}{2 m \tilde{D}_{m,n}} \frac{\Psi_{m+1,n} - 2 \, \Psi_{m,n} + \Psi_{m-1,n}}{\Delta x^2} - i \frac{\Psi_{m,n}}{\tilde{D}_{m,n}} \left( \mathcal{A}_{m,n} + \frac{E}{m} \frac{w_{m+1,n} - 2 \, w_{m,n} + w_{m-1,n}}{\Delta x^2} \right)
\end{equation}
Since far from the solitons we expect the superfluid to assume its uniform bulk value, we require the field to remain constant and its derivatives to be zero at the boundary points of the chosen spatial grid. If now, for a certain time step $t_n$, we know the values $\Psi_{m,n}$ for all positions $x_m$, the explicit RK4 method allows us to calculate for every position the value $\Psi_{m,n+1}$ of the next time step by using the following algorithm \citep{THSuliMayers}: 
\begin{align}
&p_{m,n} = f(\Psi_{m,n}) \\
&q_{m,n} = f(\Psi_{m,n}+p_{m,n}/2) \\
&r_{m,n} = f(\Psi_{m,n}+q_{m,n}/2) \\
&s_{m,n} = f(\Psi_{m,n}+r_{m,n}) \\
&\Psi_{m,n+1} = \Psi_{m,n} + \frac{\Delta t}{6}(p_{m,n} + 2 \, q_{m,n} + 2 \, r_{m,n} + s_{m,n})
\end{align}
This scheme can be repeated until the solution has been evolved up to the desired point in time. 

\subsection{Stability and convergence}
\begin{figure}[htbp]
\centering
\includegraphics[width=0.6\textwidth]{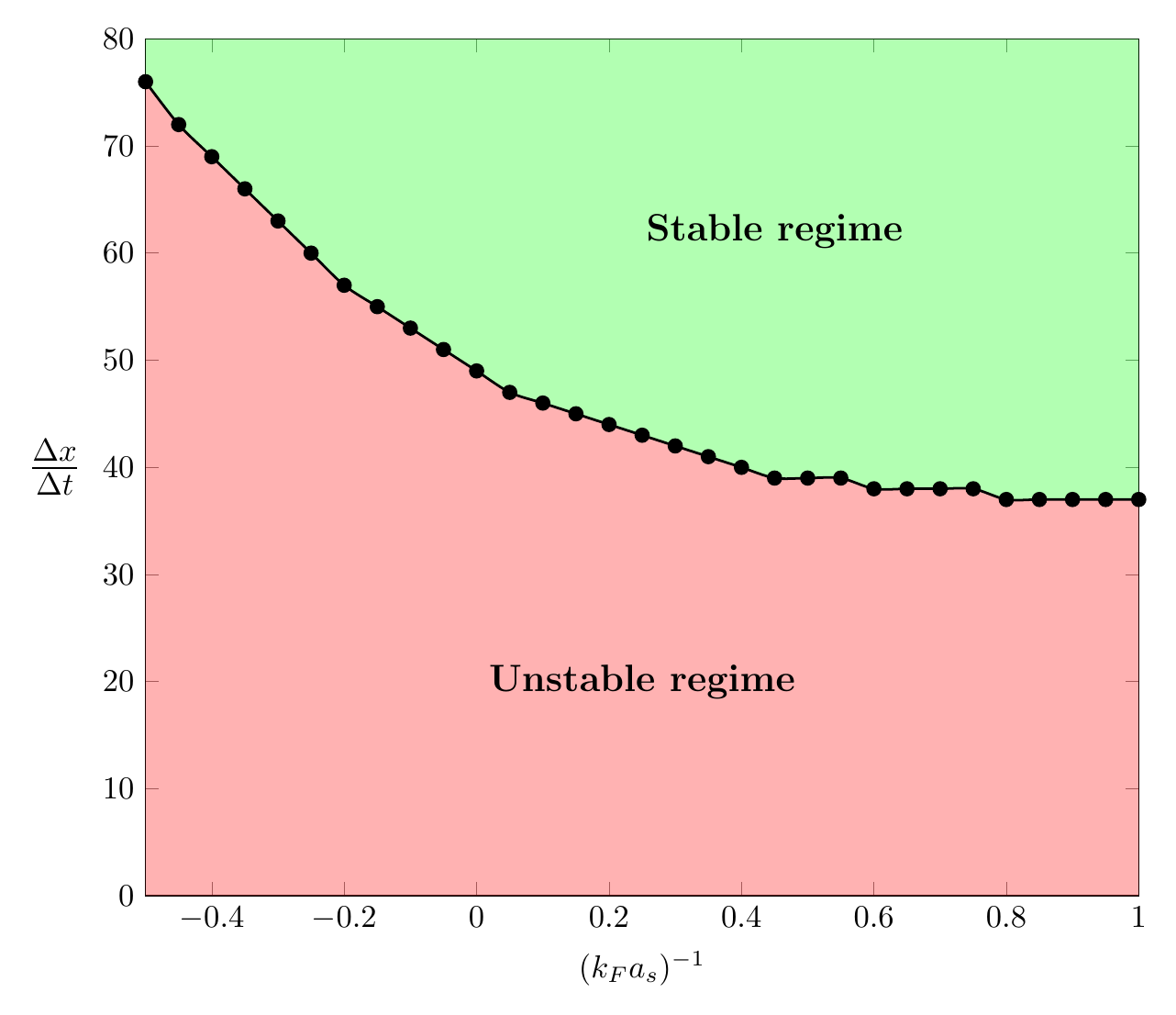}
\caption{Minimal value of the ratio $\Delta x / \Delta t$ in order to ensure stability of the numerical algorithm, in function of the interaction parameter $(k_F a_s)^{-1}$. Above the black line the algorithm is stable, while below the line it is unstable.} 
\label{fig:stab}
\end{figure}
For the finite-difference algorithm to be stable, the ratio $\Delta x / \Delta t$ must be sufficiently large, meaning that the smaller we choose $\Delta x$, the smaller the value of $\Delta t$ has to be to prevent the simulation from diverging. This results in a trade-off between the preferred resolution of the spatial grid and the amount of time steps that will be needed to reach the desired point in time. Figure \ref{fig:stab} shows the minimal value which $\Delta x / \Delta t$ must have across the interaction domain in order for the algorithm to be stable. One can observe that the condition for stability becomes stricter when moving from the BEC-regime to the BCS-regime. We find that for the grid step $\Delta x = 0.02 \, k_F^{-1}$ that is used in this work, the maximal stable values of the time step $\Delta t$ are sufficiently small to ensure that the convergence error of the finite-difference algorithm is much smaller than the error coming from the finite spatial step size. Therefore, the numerical errors on all calculated quantities are determined by the spatial resolution of our grid.

\bibliography{Refs_experiment,Refs_theory}

\end{document}